\documentclass[reprint,aps,onecolumn]{revtex4-1}
\usepackage{graphicx,epstopdf,float,setspace,mathtools}
\usepackage[usenames]{color}
\pdfoutput=1
\providecommand{\rbr}[1]{\left( #1 \right)}%
\providecommand{\ttiny}[1]{\text{\tiny{#1}}}%
\providecommand{\ttiny}[1]{\text{\tiny{#1}}}%

\begin{document}
\title{Discrete and Weyl density of states for photons and phonons}
\author{Alhun Aydin$^{1,2}$}
\author{Thomas Oikonomou$^{3}$}
\author{G. Baris Bagci$^{4}$}
\author{Altug Sisman$^{1,2}$}
 \thanks{Corresponding author: altug.sisman@physics.uu.se}
\affiliation{$^{1}$Department of Physics and Astronomy, Uppsala University, 75120, Uppsala, Sweden \\
$^{2}$Nano Energy Research Group, Energy Institute, Istanbul Technical University, 34469, Istanbul, Turkey \\
$^{3}$Department of Physics, School of Science and Technology, Nazarbayev University, Astana 010000, Kazakhstan \\
$^{4}$Department of Physics, Mersin University, 33110 Mersin, Turkey}
\date{\today}
\begin{abstract}
The current density of states (DOS) calculations do not take into account the essential discreteness of the state space, since they rely on the unbounded continuum approximation. Recently, discrete DOS based on the quantum-mechanically allowable minimum energy interval has been introduced for quadratic dispersion relation. In this work, we consider systems exhibiting linear dispersion relation, particularly photons and phonons, and calculate the related density and number of states (NOS). Also, a Weyl's conjecture-based DOS function is calculated for photons and phonons by considering the bounded continuum approach. We show that discrete DOS function reduces to expressions of bounded and unbounded continua in the appropriate limits. The fluctuations in discrete DOS completely disappear under accumulation operators. It's interesting that relative errors of NOS and DOS functions with respect to discrete ones are exactly the same as the ones for quadratic dispersion relation. Furthermore, the application of discrete and Weyl DOS for the calculation of internal energy of a photon gas is presented and importance of discrete DOS is discussed. It's shown that discrete DOS function given in this work needs to be used whenever the low energy levels of a physical system are heavily occupied.
\end{abstract}
\maketitle
\section{Introduction}
The density of states (DOS) of a system is an important concept that counts the number of states per interval of energy for each energy level \cite{Pathria}. However, the orthodox calculations of DOS in general are restricted by the unbounded continuum approximation. On the other hand, the state space is essentially discrete because of the finiteness of the domains and additionally due to the wave character of the particles. This discreteness is however overlooked by recourse to the argument that the domain sizes of interest are usually much greater than the de Broglie wavelength of the particles. This reasoning naturally allows us to make use of the continuous DOS (hereafter called CDOS) function prevalent in current literature \cite{dos1,dos2,dos3}. 

The recent progress in nano-science and nano-technology indicates that the concept of density of states is also useful in these cutting edge fields \cite{a1,a2,a3,a4,a5}. However, it should be noted that domain sizes can often be on the order of the de Broglie wavelength of particles at nano scale. When this is the case, bounded continuum approximation is more appropriate to use, since it takes into account the non-zero value of the ground states associated with the momentum components. Bounded continuum approximation forms the ingredient of the Weyl's conjecture which sets the stage to understand the behaviour of the eigenvalues asymptotically. Therefore, Weyl's conjecture provides a more accurate enumeration of the states. This we call Weyl DOS (WDOS in short). Despite this improvement though, both CDOS and WDOS have at their very cores the so-called continuum approximation, and therefore lack of capturing the feature of essential discreteness. 

The discreteness becomes particularly important when quantum confinement cannot be neglected due to the fact that quantum mechanics always implies a finite energy interval. Therefore, one should consider this quantum mechanical finiteness of the energy levels and calculate the density of states in a completely discrete manner. Recently, such an approach has been used, contributing to the limited number of studied on DOS \cite{LimStudN1,LimStudN2,LimStudN3,LimStudN4,LimStudN5,LimStudN6,LimStudN7,LimStudN8,LimStudN9}, to enumerate the number of states exactly by taking into account the finiteness of the minimum allowable energy interval as dictated by quantum mechanics, namely, the discrete density of states function (DDOS from here on) \cite{sisman}. This approach has been used for the usual but important particle in a box model and shown to lead to both bounded and unbounded continua expressions in the appropriate limits \cite{sisman}. 

The aim of this work is first to calculate the discrete density of states (DDOS) for photons/phonons (where the dispersion relation is linear) and then compare the findings with those of the usual continuous density of states (CDOS) and Weyl density of states (WDOS). We compare these results also with the ones of quadratic dispersion relation, which actually bears quite a few resemblances. Discussions and conclusion are presented in section III.                

\section{D-dimensional forms of density of states functions for linear dispersion relation}

For the particles obeying linear dispersion relation such as photons and phonons confined in a $D$-dimensional rectangular domain, dimensionless energy eigenvalues are written as
\begin{equation}\label{eq01}
\tilde{\varepsilon}=\frac{\varepsilon}{k_\ttiny{B} T}=\frac{\hbar \omega}{k_\ttiny{B} T}=\frac{hv}{2\pi k_\ttiny{B} T} \|\vec{k}\|=L_{*} \sqrt{\sum_{n=1}^{D}\rbr{\frac{i_n}{L_n}}^2}=\sqrt{\sum_{n=1}^{D}\rbr{\alpha_n i_n}^2}
\end{equation}
where $k_\ttiny{B}$ is Boltzmann's constant, $T$ is temperature, $h$ is the Planck's constant, $v$ is the speed of photons or phonons, $\vec{k}$ is wave vector, $D$ is the number of spatial dimensions, $i_n$ is quantum state variable varying from one to infinity, $L_{*}=(hv)/(2k_\ttiny{B}T)$ is a characteristic length scale, $L_n$ is the size of the domain in direction $n$ and $\alpha_n=L_{*}/L_n$ is confinement parameter in $n$'th direction. The calculations of DOS and NOS functions in this article will be done considering this dispersion. As DOS functions are used along with distribution functions in statistical mechanics to calculate physical properties of a system, throughout the article, dimensionless energy ($\tilde{\varepsilon}=\varepsilon/k_\ttiny{B} T$) is adopted rather than the energy itself for the compactness of the expressions.

DDOS function converts multiple summations over discrete quantum states into a single summation over discrete energy states as follows \cite{sisman}
\begin{equation}\label{eq02}
\sum_{i_1=1}^{\infty} \cdots \sum_{i_D=1}^{\infty} f(\tilde{\varepsilon}_{i_1},\cdots, \tilde{\varepsilon}_{i_D})\Delta i_1\cdots \Delta i_D =\sum_{\tilde{\varepsilon}=\tilde{\varepsilon}_0}^{\infty} f(\tilde{\varepsilon})\;\textit{DDOS}(\tilde{\varepsilon}) \;\Delta\tilde{\varepsilon}
\end{equation}
where $\tilde{\varepsilon}_0=\sqrt{\alpha_1^2+\cdots+\alpha_D^2}$ is the ground state energy and $\Delta\tilde{\varepsilon}$ is quantum-mechanically minimum allowed energy difference between successive levels. Note that $\Delta\tilde{\varepsilon}$ is not a constant quantity unlike $d\tilde{\varepsilon}$. The analytical solution of $\Delta\tilde{\varepsilon}$ is only possible for 1D case so that the numerical calculation of the energy intervals between successive levels requires generating the energy spectrum data through Eq. (1) first and then applying the ascending sorting process. Hence, the definition of DDOS is given by 
\begin{equation}\label{eq03}
\textit{DDOS}_{D}(\tilde{\varepsilon})=\frac{\Delta\Omega_D(\tilde{\varepsilon})}{\Delta\tilde{\varepsilon}}=\frac{\Omega_{D}(\tilde{\varepsilon}+\Delta\tilde{\varepsilon})-\Omega_{D}(\tilde{\varepsilon})}{\Delta\tilde{\varepsilon}}
\end{equation}
where $\Omega_D$ is discrete number of states (DNOS) given by,
\begin{equation}\label{eq04}
\Omega_D(\tilde{\varepsilon})=\textit{DNOS}_{D}(\tilde{\varepsilon})=\sum_{i_1^{\prime}=1}^{\infty}\cdots\sum_{i_D^{\prime}=1}^{\infty}\Theta\left[\tilde{\varepsilon}-\sqrt{\sum_{n=1}^{D}{(\alpha_n i_n^{\prime})^2}}\right]
\end{equation}
where $\Theta$ is left-continuous Heaviside step function so that $\Theta(0)=0$.

Considering Weyl's conjecture in a $3$-dimensional finite-size domain \cite{Pathria}, it is possible to write the WNOS function using the dispersion relation of Eq. (1) as 
\begin{equation}\label{eq05}
\begin{split}
\textit{WNOS}_D(\tilde{\varepsilon}) =&
\frac{\pi}{6}\frac{V}{L_{*}^{3}}\tilde{\varepsilon}^{\,3}\;\Theta(D-2) 
+ (-1)^D \frac{\pi}{4^{D-2}4}\frac{S}{L_{*}^{2}}\tilde{\varepsilon}^2\;\Theta(D-1) \\
& + (-1)^{D-1}\frac{1}{4^{D-1}}\frac{P}{L_{*}}\tilde{\varepsilon}\;\Theta(D) + (-1)^{D-2}\frac{N_V}{4^D}
\end{split}
\end{equation}
and from the derivative of WNOS, we get WDOS function as
\begin{equation}\label{eq06}
\begin{split}
\textit{WDOS}_D(\tilde{\varepsilon}) = &
\frac{\pi}{2}\frac{V}{L_{*}^{3}}\tilde{\varepsilon}^2\;\Theta(D-2)
+ (-1)^D \frac{\pi}{4^{D-2}2}\frac{S}{L_{*}^{2}}\tilde{\varepsilon}\;\Theta(D-1) \\
& + (-1)^{D-1}\frac{1}{4^{D-1}}\frac{P}{L_{*}}\Theta(D).
\end{split}
\end{equation}
where $V$ is volume, $S$ is surface area, $P$ is periphery and $N_V$ is the number of vertices. These parameters together constitute the geometric size variables of the domain. Note that Eqs. (5) and (6) are not just for rectangular geometries but universal so that they are valid for any confinement geometry. Using Eqs. (5) and (6), neglecting the high order correction terms (terms having low dimensional size variables), CNOS and CDOS functions can be obtained for $D$-dimensional rectangular domain as
\begin{equation}\label{eq07}
\textit{CNOS}_{D}(\tilde{\varepsilon})=\frac{\pi^{D/2}\tilde{\varepsilon}^D L_1\cdots L_{{}_{D}}}{2^D\Gamma[(D+2)/2]L_{*}^{{}^{D}}}
\end{equation}
\begin{equation}\label{eq08}
\textit{CDOS}_{D}(\tilde{\varepsilon})=\frac{\pi^{D/2}\tilde{\varepsilon}^{D-1}L_1\cdots L_{{}_{D}}}{2^{D-1} \Gamma[D/2]L_{*}^{{}^{D}}},
\end{equation}
respectively.

The subbands appear for confined directions in the case of lower dimensional structures. To this aim, $\tilde{\varepsilon}$ variables in the equations of this article need to be modified to $\sqrt{\tilde{\varepsilon}^2-\tilde{\varepsilon}_s^2}$ for lower dimensions and the summation operator $\sum_{\tilde{\varepsilon}_s}\Theta\left(\sqrt{\tilde{\varepsilon}^2-\tilde{\varepsilon}_s^2}\right)\times\cdots$ over subbands must be added to the expressions.

\subsection{Analytical expressions of WNOS and WDOS functions in various dimensions}
Analytical expressions of WDOS functions are obtained by considering bounded continuum approximation, which corresponds to the first two terms of Poisson Summation Formula (PSF) \cite{Spiegel}. To find the analytical expressions of WNOS and WDOS functions for linear dispersion relation, we first apply the same methodology, based on the first two terms of PSF, presented in \cite{sisman} to DNOS function, Eq. (4). Note that the derivations for lower dimensional cases are carried out without considering the subbands of confined directions, so the 2D and 1D treatments are identified as if the actual dimension of the space is not 3D but lower dimensional.

By replacing summations in Eq. (4) by the first two terms of PSF, WNOS functions for 3D, 2D and 1D cases are obtained respectively as
\begin{subequations}\label{eq9}
\begin{align}
& \textit{WNOS}_3= \frac{\pi\tilde{\varepsilon}^{3}}{6\alpha_1\alpha_2\alpha_3}-\frac{\pi\tilde{\varepsilon}^2}{8}\left(\frac{1}{\alpha_1\alpha_2}+\frac{1}{\alpha_1\alpha_3}+\frac{1}{\alpha_2\alpha_3}\right)+\frac{\tilde{\varepsilon}}{4}\left(\frac{1}{\alpha_1}+\frac{1}{\alpha_2}+\frac{1}{\alpha_3}\right)-\frac{1}{8}, \\
& \textit{WNOS}_2=\frac{\pi\tilde{\varepsilon}^2}{4\alpha_1\alpha_2} - \frac{\tilde{\varepsilon}}{2}\left(\frac{1}{\alpha_1} + \frac{1}{\alpha_2}\right) + \frac{1}{4}, \\
& \textit{WNOS}_1=\frac{\tilde{\varepsilon}}{\alpha_1}-\frac{1}{2}.
\end{align}
\end{subequations}

The first terms of each equation above gives the corresponding CNOS functions. Straightforward differentiation of the above equation yields
\begin{subequations}\label{eq10}
\begin{align}
& \textit{WDOS}_3=\frac{\pi\tilde{\varepsilon}^2}{2\alpha_1\alpha_2\alpha_3}-\frac{\pi\tilde{\varepsilon}}{4} \left(\frac{1}{\alpha_1\alpha_2}+\frac{1}{\alpha_1\alpha_3}+\frac{1}{\alpha_2\alpha_3}\right)+\frac{1}{4}\left(\frac{1}{\alpha_1}+\frac{1}{\alpha_2}+\frac{1}{\alpha_3}\right), \\
& \textit{WDOS}_2=\frac{\pi\tilde{\varepsilon}}{2\alpha_1\alpha_2}-\frac{1}{2}\left(\frac{1}{\alpha_1}+\frac{1}{\alpha_2}\right), \\
& \textit{WDOS}_1=\frac{1}{\alpha_1}.
\end{align}
\end{subequations}
The first terms of each equation above gives the corresponding CDOS functions. It is appropriate to use CNOS and CDOS functions only if confinement parameters are much smaller than unity.

It should be noted that exact analytical expression of DDOS is equal to WDOS for 1D case unlike the 2D and 3D cases, since only one state is present in each energy level i.e. $\Delta\Omega_1=\Omega_{1}(\tilde{\varepsilon}+\Delta\tilde{\varepsilon})-\Omega_{1}(\tilde{\varepsilon})=\Delta i=1$. Due to the lack of degeneracy, we simply have  $\Delta\tilde{\varepsilon}=\alpha_1$. As a result, DDOS can analytically be obtained as
\begin{equation}\label{eq11}
\textit{DDOS}_1=\frac{\Delta\Omega_1}{\Delta\tilde{\varepsilon}}=\frac{1}{\alpha_1}=\textit{WDOS}_1=\textit{CDOS}_1
\end{equation}
only for 1D case.

\section{Results and Discussion}

\subsection{Comparisons of discrete, Weyl and classical DOS and NOS functions}
Analytical expressions of WNOS and WDOS as well as CNOS and CDOS functions are given without considering subbands. During the calculations of figures for lower dimensional cases, however, the contributions of subbands are considered by using the modification procedures defined for the related equations in the first paragraph after Eq. (8). Thus, during the examination of functional behaviors of DOS and NOS functions, $\tilde{\varepsilon}$ is replaced by $\sqrt{\tilde{\varepsilon}^2-\tilde{\varepsilon}_s^2}$ in all expressions to see the effects of the sub-bands. In Fig. 1(a, b, c), Eqs. (3), (10) and (8) are used to draw DDOS, WDOS and CDOS functions represented by black dots, blue and red curves, respectively. The extremely fluctuating feature of DDOS is evident in Fig. 1(a, b, c) when compared with the rather smooth behaviour of CDOS and WDOS. In this sense, DDOS is essentially seen to be very different than both CDOS and WDOS. However, the usual thermodynamic practice almost always includes DOS functions to be smoothed out through a summation or integration process. If this was not the case, it is apparent that the use of DOS does not yield the same results as CDOS or WDOS. The transition from this extremely fluctuating behaviour of DDOS to a rather smooth one can be explicitly observed by inspecting Fig. 1(d, e, f) and comparing the corresponding DDOS and DNOS plots. In Fig. 1(d, e, f), Eqs. (4), (9) and (7) are used to draw DNOS, WNOS and CNOS functions represented by black dots, blue curves and red curves respectively. Note that WNOS and CNOS too are different from one another as can be seen from Fig. 1(d, e, f), although they both rely on the continuum approximation. The difference lies in the fact that WNOS makes use of the bounded continuum approximation whereas CNOS of the unbounded one. Since WNOS almost perfectly matches with DNOS, it should be favored to CNOS as long as confinement is not negligibly weak. 

\begin{figure}[h]
\centering
\includegraphics[width=0.75\textwidth]{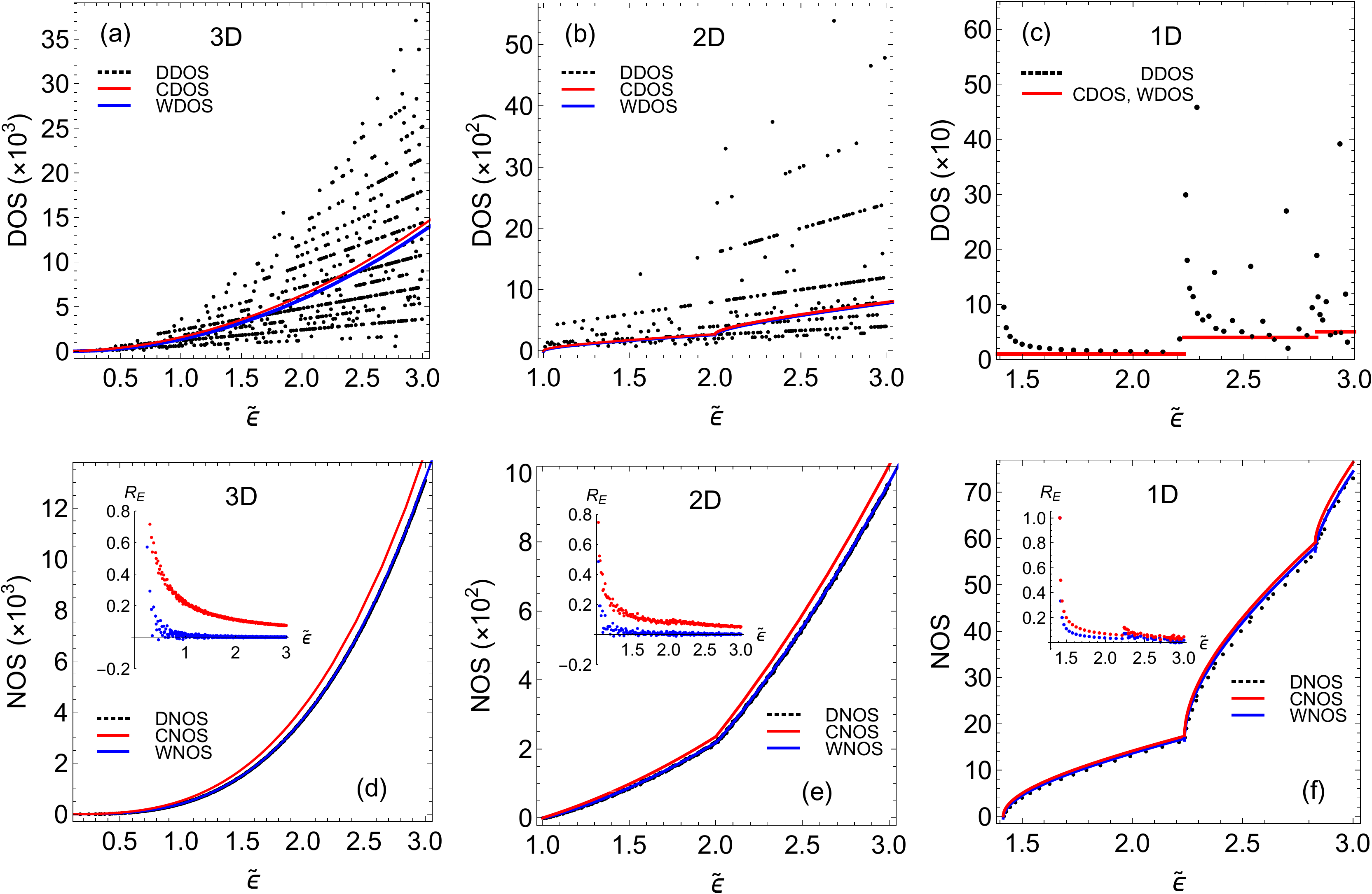}
\caption{DDOS, CDOS and WDOS functions (upper row), DNOS, CNOS and WNOS functions and its relative errors (lower row) varying with energy. (a), (d) 3D box with $\alpha_1=\alpha_2=\alpha_3=0.1$, (b), (e) Quasi-2D quantum well with $\alpha_1=\alpha_2=0.1, \alpha_3=1$ and (c), (f) Quasi-1D quantum wire with $\alpha_1=0.1, \alpha_2=\alpha_3=1$.}
\label{fig:pic1}
\end{figure}

When DOS and NOS functions given in Fig 1. are compared with the ones given in Ref. \cite{sisman} for quadratic dispersion relation, it is seen that the dimensional characteristic behaviors changed. For example, constant-like nature of 2D DOS function for quadratic dispersion is seen in 1D DOS here, or square root dependence of 3D DOS function on energy for quadratic dispersion is seen here in 2D DOS. Similar change in the functional dependence on energy can be seen also in NOS functions. Functional energy dependence can easily be predicted from the powers of $\tilde{\varepsilon}$ in Eqs. (9) and (10). Although 1D NOS function seems to have a linear dependency on energy, Eq. (9c), this dependence only appears in a hypothetical 1D world where all subbands are disappeared. On the other hand, in reality 1D behavior appears when other two directions are strongly confined and constitute subbands. Therefore, due to the existence of subbands, even for 1D case non-linear behavior appears at the beginning of each subband since $\tilde{\varepsilon}$ is replaced by $\sqrt{\tilde{\varepsilon}^2-\tilde{\varepsilon}_s^2}$, Fig. 1f. Moreover, subfigures of Fig. 1(d, e, f) shows that errors of WNOS functions always lower than CNOS, while errors of both functions increase with decreasing energy near ground state.

The gradual changes of relative errors with increasing confinement and energy as well as with increasing NOS are examined in three-dimensional graphs in Fig. 2a and 2b. It is seen that relative errors oscillates around zero (having both positive and negative values), after starting from very large error values at near ground state energies or low NOS. Increment in confinement shifts the ground state energy to higher energies and grows the magnitude of oscillations. As expected, errors decrease with increasing energy and NOS or decreasing confinement. Relative errors decrease with increasing NOS just like in the quadratic dispersion case. In fact, it is worth to emphasize that relative error results are not specific for the linear dispersion relation, the variation of relative errors is exactly the same for quadratic dispersion relation also (see Ref. \cite{sisman}). Since linear dispersion relation considered in this article is just the square root of the energy in quadratic one, the relative errors become the same for both dispersions while DDOS and DNOS functions are different.

\begin{figure}[h]
\centering
\includegraphics[width=0.7\textwidth]{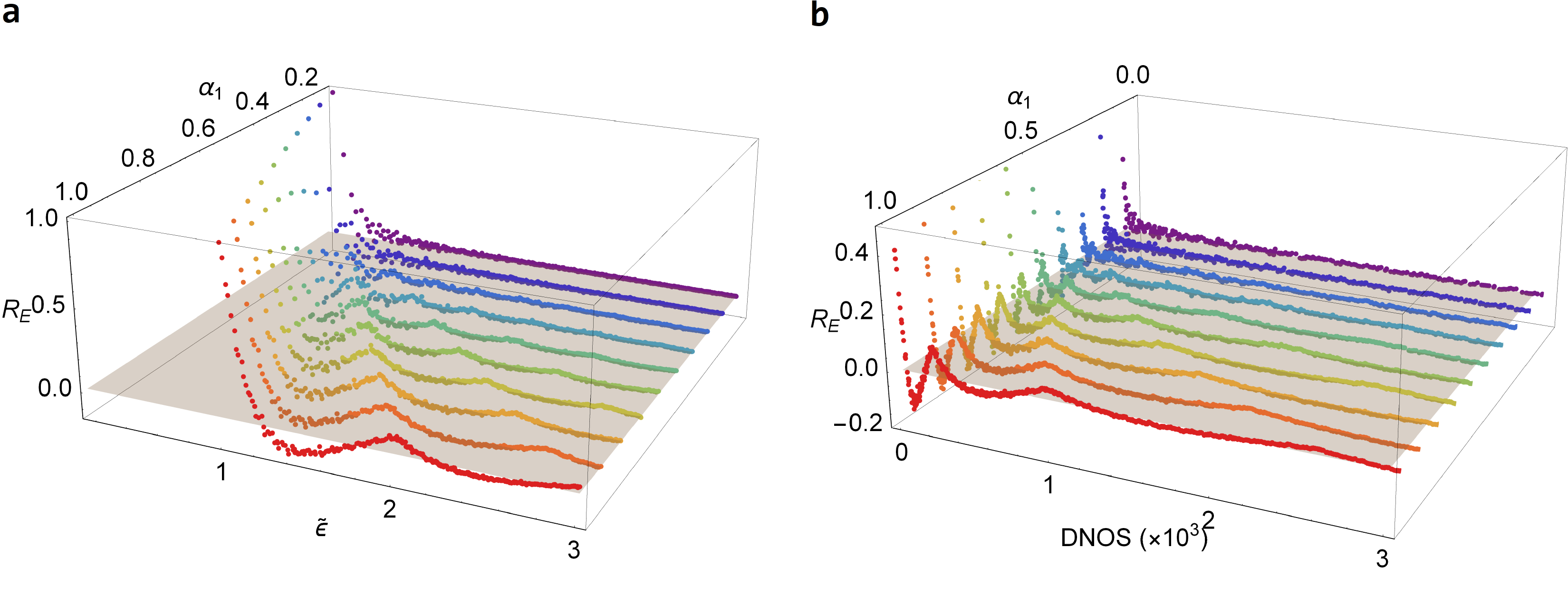}
\caption{Relative errors of WNOS for 10 different confinement values ranging from $\alpha_1=0.1$ to $\alpha_1=1$ varying with (a) energy (b) DNOS. For all cases other two directions are taken as nearly free so that $\alpha_2=\alpha_3=0.1$.}
\label{fig:pic2}
\end{figure}

\subsection{Internal energy of a confined photon gas: Comparison of the results based on different DOS functions}

From the relative errors of NOS functions in previous subsection, it's seen that WNOS very accurately represents the exact behavior as long as the system is not very close to the ground state. In this subsection, we show the accuracy and limitations of WDOS function through examining the internal energy of a photon gas that is an important quantity also for the calculation of micro/nanoscale radiative heat transfer for instance. We calculate internal energy of a photon gas confined in a rectangular domain by using discrete, Weyl and continuous DOS functions. In Fig. 3, comparisons of internal energy values calculated from DDOS, WDOS and CDOS functions denoted by black, dashed-cyan and orange curves respectively are shown for various dimensional isometric and anisometric confinement cases. It is seen that using CDOS functions in internal energy calculations give inaccurate results especially in case the system is strongly confined at least in one direction (Fig. 3d). On the other hand, internal energy calculated using WDOS functions quite fairly matches with the exact results based on DDOS. However, as is seen from the relative errors of Weyl internal energies (blue curves) in subfigures of Fig. 3(a, b, c, d), even using WDOS gives incorrect results when the confinement is strong enough, though internal energy itself has relatively small values in this condition. Nevertheless, this condition can appear at very low temperatures or in very small confinement domains.

\begin{figure}[h]
\centering
\includegraphics[width=0.95\textwidth]{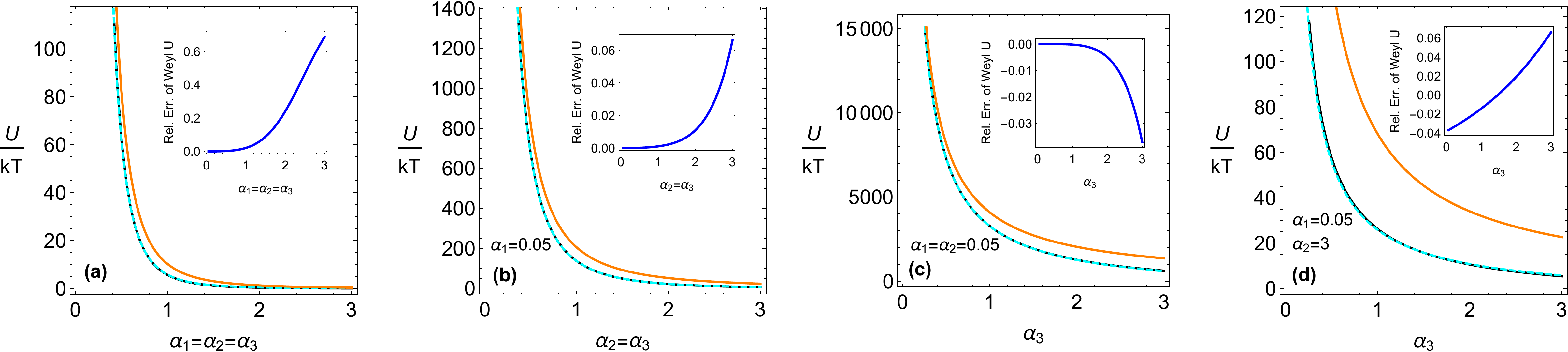}
\caption{Dimensionless internal energy changing with different confinement values. Black, dashed-cyan and orange curves represent internal energy values determined by discrete, Weyl and continuous DOS functions respectively. (a) Isometric confinement ($\alpha_1=\alpha_2=\alpha_3$ changes), (b) 1D isometric ($\alpha_1=0.05$, $\alpha_2=\alpha_3$ changes), (c) 2D isometric ($\alpha_1=\alpha_2=0.05$, $\alpha_3$ changes), (d) anisometric ($\alpha_1=0.05$, $\alpha_2=3$, $\alpha_3$ changes). Subfigures represent the relative error of internal energy calculated from WDOS for each particular case.}
\label{fig:pic3}
\end{figure}

\subsection{Conclusion}
To sum up, when the low energy levels are largely occupied as in the cases of low temperature or strong confinement, DDOS gives better results than even WDOS. However, since the calculation of DDOS requires more computational efforts than the usual DOS, it is reasonable to adopt WDOS function in most of the cases, since it yields relatively small (compared to CDOS) errors. In other words, WDOS function gives much more accurate representation for the true behavior of DOS function than CDOS for almost all ranges except the energies near to ground state energy. Hence, WDOS functions should be used in the calculation of physical properties of many systems ranging from nano to micro scale. On the other hand, DDOS may need to be used in the examination of high confinement cases or systems having the large portion of particles near to ground state as near ground state properties can be more accurately predicted by using DDOS function.

\section*{Acknowledgements}
The author A.A. gratefully acknowledges support from AIM Energy Technologies A.S. The author T.O. acknowledges the state-targeted program "Center of Excellence for Fundamental and Applied Physics" (BR05236454) by the Ministry of Education and Science of the Republic of Kazakhstan and ORAU grant entitled "Casimir light as a probe of vacuum fluctuation simplification" with PN 17098.


\end{document}